\documentclass[aip,jcp,amsmath,amssymb,reprint,superscriptaddress,groupedaddress]{revtex4-1}
\usepackage{mathtools}
\usepackage{amsthm}
\usepackage{amssymb}
\usepackage{bm}
\usepackage{expl3}
\usepackage{anyfontsize}
\usepackage{braket}
\usepackage{relsize}
\usepackage{enumitem}
\usepackage[vlined]{algorithm2e}
\usepackage{algpseudocode}
\usepackage[compact]{titlesec}
\usepackage{ragged2e}
\usepackage{caption}
\usepackage{graphicx}
\usepackage{footnote}
\usepackage{multirow}
\usepackage{tabularx}
\usepackage{booktabs}
\usepackage{makecell}
\usepackage{threeparttable}
\usepackage{color}
\usepackage{breakurl}

\usepackage{float}

\newcolumntype{C}{>{\centering\arraybackslash}X}

\begin{document}

\title{Learning to Make Chemical Predictions: the Interplay of Feature Representation, Data, and Machine Learning Algorithms}

\author{Mojtaba Haghighatlari$^{1*}$, Jie Li$^{1*}$,Farnaz Heidar-Zadeh$^{1-3*}$, Yuchen Liu$^{1}$,Xingyi Guan$^{1,4}$, Teresa Head-Gordon$^{1,4-5}$}
\affiliation{$^1$Kenneth S. Pitzer Theory Center and Department of Chemistry, University of California, Berkeley, CA, USA, *all authors contributed equally}
\affiliation{$^2$Center for Molecular Modeling (CMM), Ghent University, B-9052 Ghent, Belgium}
\affiliation{$^3$Department of Chemistry, Queen's University, Kingston, Ontario K7L 3N6, Canada}
\affiliation{$^4$Chemical Sciences Division, Lawrence Berkeley National Laboratory, Berkeley, CA, USA}
\affiliation{$^5$Departments of Bioengineering and Chemical and Biomolecular Engineering, University of California, Berkeley, CA, USA,  corresponding author:thg@berkeley.edu}

\begin{abstract}
Recently supervised machine learning has been ascending in providing new predictive approaches for chemical, biological and materials sciences applications. In this Perspective we focus on the interplay of machine learning algorithm with the chemically motivated descriptors and the size and type of data sets needed for molecular property prediction. Using Nuclear Magnetic Resonance chemical shift prediction as an example, we demonstrate that success is predicated on the choice of feature extracted or real-space representations of chemical structures, whether the molecular property data is abundant and/or experimentally or computationally derived, and how these together will influence the correct choice of popular machine learning algorithms drawn from deep learning, random forests, or kernel methods. 
\end{abstract}

\maketitle


\begin{figure}[H]
\center
\includegraphics[width=7 cm]{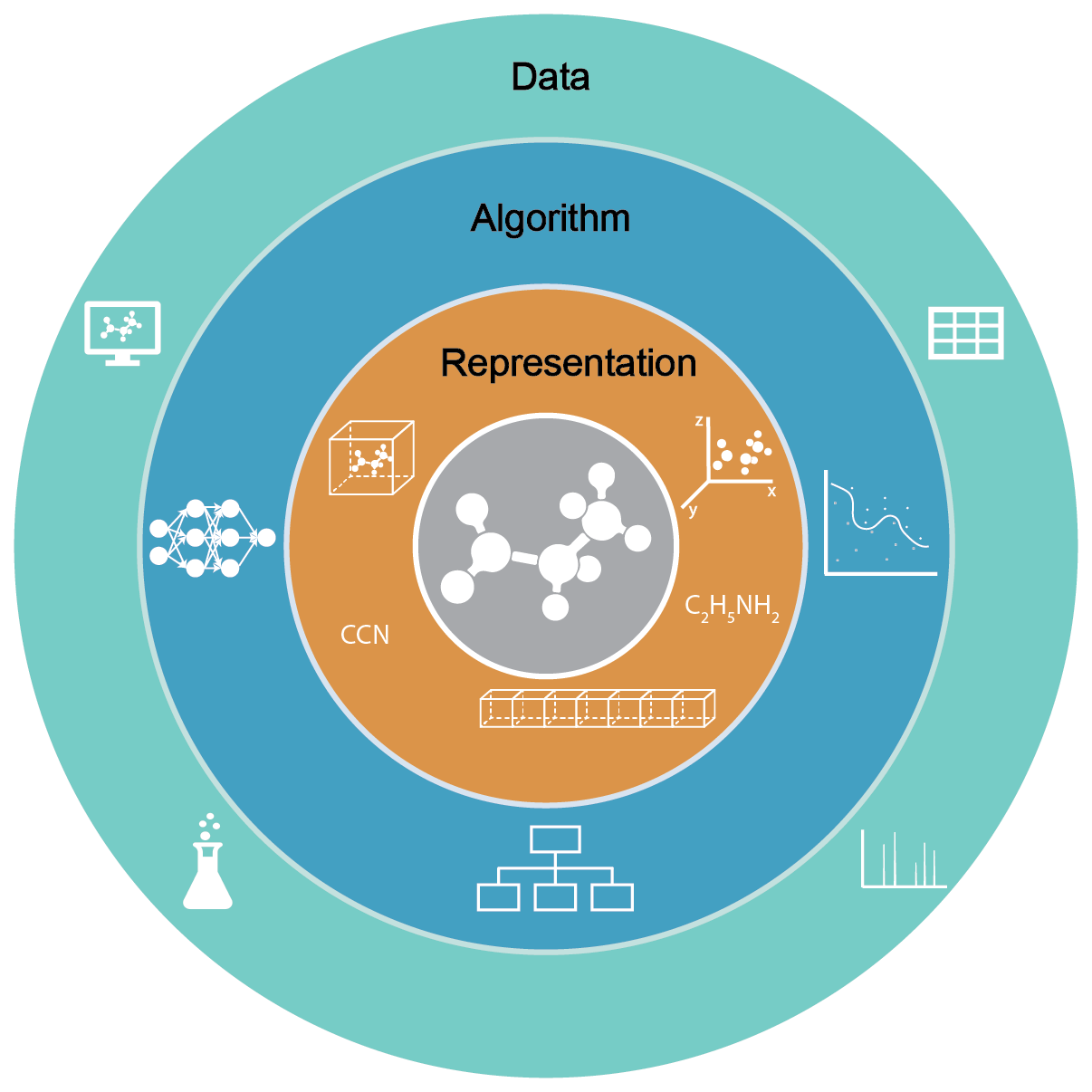}
\caption*{\fontsize{8}{8}\selectfont Success of chemical prediction with machine learning relies heavily on the interplay of three key components: representation, algorithm and data. Molecules can be represented as extracted descriptors (fingerprints, etc.) or direct real-space representations (for example 3D coordinates or densities). Common algorithms for input-output mapping including deep learning models, kernel methods and random forests are of different model complexity. Type and quality of data must be considered to achieve an effective integration of the three aspects.}
\end{figure}

\section*{\fontsize{11}{11}\selectfont INTRODUCTION}
\label{sec:intro}
\noindent
The rise (again) of machine learning (ML) in the molecular sciences is a  transformation of the traditional ways in which we perform computational chemistry. Unlike von Neumann machine algorithms, which articulate a mathematical model whose equations can be solved in a logical progression, machine \textit{learning} computational chemistry is formulated as "non-algorithmic" computing using (typically) supervised learning of well-curated data to map molecules to chemical properties. With appropriate strategies, ML has been successfully applied to quantum mechanically derived energy and force evaluation\cite{Chmiela2019, Smith2019a, Amabilino2019}, molecular dynamics\cite{Wang2020a}, three-dimensional structure prediction of small molecule crystals to large proteins\cite{Sanchez-Lengeling2018,alphafold_nature, alquraishi2019RGN}, pathways for chemical reactivity and catalysis \cite{Brickel2019a, Shakouri2017, Schwaller2019}, and the rapid evaluation of spectroscopic and molecular properties\cite{ucbshift_arxiv,3d_densenet,Yang2019,Haghighatlari2019d}.

ML has a long and storied history that builds on traditional mathematical programming and statistical and clustering models, and early meta-heuristic methods such as genetic algorithms and artificial neural networks (ANNs)\cite{Russell2010}. Broadly speaking, the most popular machine learning algorithms used in the chemical sciences today have evolved from these early efforts to now include non-parametric statistical learning such as decision trees and random forests, kernel-based models such as Gaussian Process regression (GPR) or Kernel Ridge regression (KRR), and deep learning (DL) networks exemplified by convolutional neural networks (CNNs)\cite{Goodfellow-et-al-2016}. 

Although machine learning algorithms were developed primarily by statisticians or computer scientists for other tasks such as image recognition\cite{LeCun1989}, the chemical sciences domain has arguably advanced most effectively the development of novel feature representations, or \textit{descriptors}, that informs the physical nature of the input-output mapping. These well designed descriptors offer many benefits including greater interpretability of the ML approach, to incorporate physical constraints on the learning parameters, or to better utilize a ML surrogate model for classification or regression. 

But in order for ML algorithms and chemical  descriptors to be effective requires the appropriate form and amount of the training data. If there is abundant training data which covers a wide scope of chemical space, it empowers DL networks with their (typically) huge number of parameters to discover complicated patterns in the data through successive transformations through their layers. For example, the popular CNNs have utilized widely available 3D representations in successful application to enzyme classification \cite{cnn_enzyme}, molecular representations \cite{cnn_rep}, and amino acid environment similarity analysis \cite{cnn_aa}. On the other hand, small datasets with a well formulated chemical representation can still be utilized by statistical or kernel models to make faithful predictions, such as predicting electronic structure correlation energies using sparse Hartree–Fock input \cite{Welborn2018}. Hence the choice of machine learning approach will be decided by whether the data stems from first-principles but limited in quantity due to expensive calculations from quantum mechanics (QM) or from abundant inexpensive calculations, or experimental data that may also be noisy, error prone, or difficult to interpret.

In this perspective, we first describe the three elements of successful prediction: ML algorithms, chemical feature representations, and dataset sizes and quality. We then illustrate their interplay for predicting  nuclear magnetic resonance (NMR) chemical shifts, either through a combination of engineered features with random forest regression for protein NMR chemical shifts in solution\cite{ucbshift_arxiv} compared to shallow ANNs, while a deep learning CNN can improve performance over a KRR for chemical shift prediction in the solid state by exploiting physically motivated data augmentation\cite{3d_densenet}. Finally we conclude with an outlook for future directions of machine learning in the areas of feature representation development, data scarcity and sparsity, as well as physics-infused models and approaches to greater interpretability of machine learning. 
\newline
\section*{\fontsize{10}{11}\selectfont THE COMPONENTS OF MACHINE LEARNING}
\vspace{6 pt}
\subsection*{\fontsize{10}{11}\selectfont Popular ML Algorithms in the Chemical Sciences}
\noindent
Artificial neural network methods attempt to map the input-output relationship through a mathematical model which resembles the connections of neurons in a mammalian brain. In the chemical context, the input of a supervised machine learning model is a  "representation", $\textbf{x}$, of a group of atoms that may form a drug molecule, a protein, a crystal structure, etc, and the output, $\textbf{y}$, is the chemical property of interest. 

The most basic computing element of an ANN, the simple perceptron\cite{Rosenblatt1958}, is capable of performing linear or logistical regression and classification with appropriate activation functions (Figure 1), and can perform Boolean operations such as the simple OR and AND functions. A slightly more complex architecture is needed when executing the exclusive XOR function that requires a pre-processing "hidden" layer between the input and output layers to appropriately define the linear decision boundaries that separates its solution space. Such early shallow ANN architectures, using everything from hand-crafted features to molecular structures, have successfully predicted more than 20 different types of physiochemical properties of a molecule, such as water solubility, Henry's law constant, heats of formation and crystal packing.\cite{physicalchemical_properties}

The universal approximation theorem states that a single hidden layer with many simple perceptrons and suitable activation functions can represent any function of \{$\textbf{x}$\} to predict $f(y|\{\textbf{x}\})$, regardless of complexity or how non-linear is its solution space.  However what is not guaranteed is that there is a universal procedure for how to \textit{learn} the transformation \{$\textbf{x}$\}$\rightarrow$ $f(y|\{\textbf{x}\})$
using a single layer architecture, nor what is the best feature representation of \{$\textbf{x}$\} to ensure that it will perform well on previously unobserved target function data. Hence most of the recent excitement in machine learning is the emergence of DL architectures, a meta-heuristic approach that replaces a single hidden layer with many, many hidden layers each composed of many artificial neurons. 
The DL network learns the input-output representations by minimizing a loss function through adjustments of the weights that connect the neuronal nodes of its architecture. 

\begin{figure}[H]
\center
\includegraphics[width=0.48\textwidth]{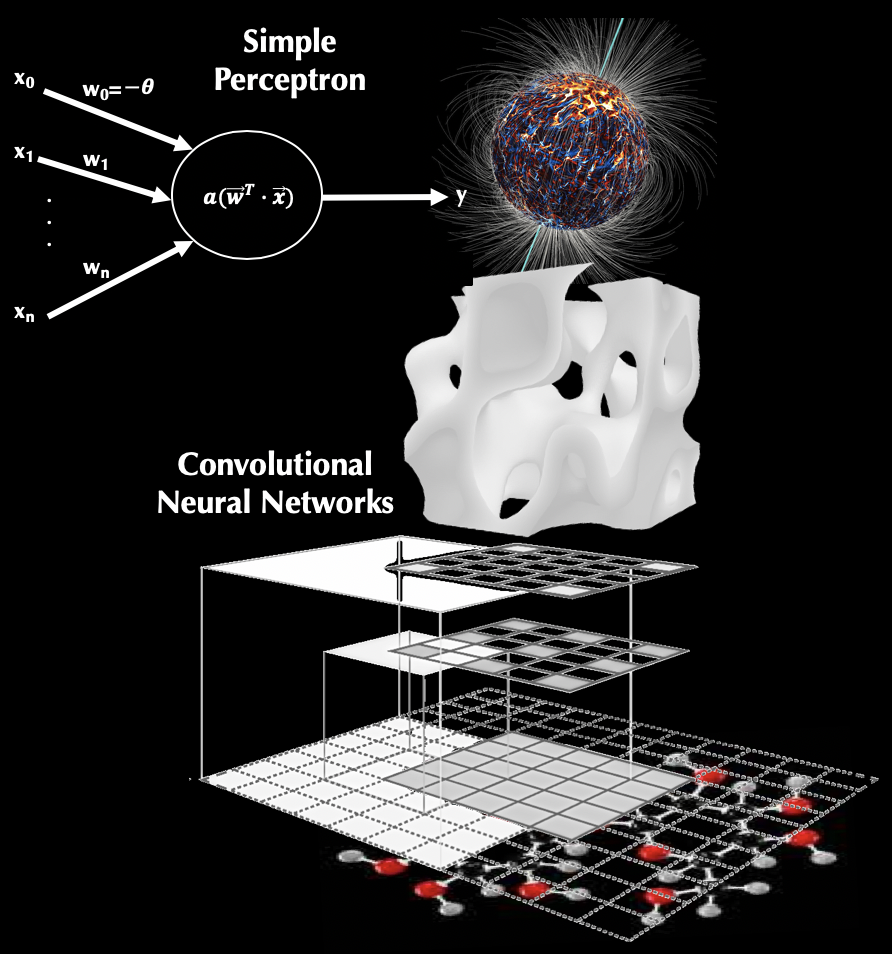}
\caption*{Figure 1. The simple perceptron of an ANN compared to the transformation of a 2D representation of a molecule with convolutions accumulated through layers of a CNN to yield atomic magnetic properties in a molecular framework, such as a chemical shift or scalar coupling value.}
\label{fig:CNNChemShift}
\end{figure}

The most classical example of a DL architecture are the CNNs that were originally introduced and popularized by LeCun for handwriting and other image recognition tasks\cite{LeCun1989}. CNNs are neural networks that use convolution operations in place of general matrix multiplication (as in standard ANNs) in at least one of their layers. During the learning process the convolutional layers typically generate multiple feature maps that when aggregated together represent new formulations of the input data. Figure 1 pictorially displays how the input data is "transformed" by the processing units of the convolution through many layers. In order to aid the learning strategy of a CNN, the sparser L connections between L convolutional layers have been recently replaced by a "denser" network of L(L+1)/2 direct connections, also known as a "DenseNet"\cite{Huang2019}. In this case the feature maps of all preceding layers are used as inputs to a current convolution layer, and its own resulting
feature maps are then used as inputs into all subsequent layers of the deep layered architecture. 

The primary distinction of a DL architecture is its much greater network capacity relative to early ANN's, and thus its greater advantage in handling much larger data sets than previously possible. The DL approach has advanced through better learning heuristics that are now well established\cite{Goodfellow-et-al-2016}:  regularization through appropriate loss functions and back-propagation, data augmentation using noise injection or non-linear transformations, and the use of dropout and batch normalization; adaptive learning strategies that bear strong equivalence to a Newton step using preconditioners that are combined with stochasticity in the gradients as per methods like RMSProp\cite{rmsprop} and Adam\cite{adam}; and finally the finetuning of the "hyperparameters" in all of these learning choices through formulations of validation data sets and through methods such as early stopping and ensemble prediction.

As such, DL is ready for prime time in the chemical sciences as their architectures can be adapted to many types of problems, their hidden layers reduce the need for feature engineering, and they have benefited from  several important regularizations that allows them to efficiently learn from high-dimensional data. At the same time DL approaches are not always suitable as general-purpose ML algorithms because they have orders of magnitudes more parameters to estimate, they require much more expertise to tune (i.e. to set the architecture and optimize the hyperparameters), and especially because they require a very large amount of well-curated labelled data. 

Alternatively, machine learning methods such as GPR and KRR can be traced back to the advent of Support Vector Machines (SVMs), which were the first machine learning algorithm with a solid mathematical foundation\cite{Russell2010}. These kernel-based methods are formulated to capture the similarities of a collection of data points through a clever choice of the "kernel". If the optimal kernel is found, the simplest linear regression is sufficient to predict the target value from its input data using  similarity to the input features of the training dataset. As such kernel methods are  powerful supervised classifiers  that optimize non-linear decision boundaries directly. They have been found to be superior to multiple linear regression and radial basis function neural networks when applied to chemical toxicity prediction for example\cite{svm_toxic}.
More recently, KRR has realized excellent performance on regression prediction for molecular properties such as NMR chemical shifts for small molecules either in solution\cite{rupp2015machine,gerrard2020impression} or in the solid state\cite{ml_nmr2}. In this case the physical understanding of a chemical system helped in the creation of a reasonable kernel function. Specifically the SOAP kernel\cite{Bartok2013c} is explicitly designed to faithfully represent an atomic environment of a molecule with uniqueness. Furthermore kernel methods naturally incorporate  symmetry functions for which it is often desirable to enforce translational or rotational invariances that may be relevent to the chemical prediction\cite{Bartok2013c,Behler2007}.

While kernel methods work very well in practice, and are robust against overfitting even in high-dimensions, they are tricky to tune due to the importance of picking the right kernel, and if the kernel function is not smooth enough in the space of the atomic environment, the resulting kernel-based method will suffer from  outliers in the training dataset that will degrade prediction performance. They also require the storage of and operation on all of the support or feature vectors, which can be prohibitive for application to large datasets. Especially in the case of KRR and GPR, because the similarity kernel needs to be applied between the pairwise features with all data examples in the training dataset, its unfavorable scaling with the number of training examples prevents it from benefiting from large datasets. 

Often statistical models such as decision trees are preferred over kernel methods as they are more robust to outliers, are much more computationally scalable, and do not require the luck of finding the kernel function as they quite naturally model non-linear decision boundaries thanks to their hierarchical structure.\cite{Goodfellow-et-al-2016} In a statistical learning model such as decision trees, training comprises the optimal splitting of the features driven by a decrease in the maximum entropy loss function from information theory. Decision tree models are equally suited for big or small datasets because once the cutting points have been identified, the application of the algorithm to new data is just a constant of time. The classification or regression prediction from a statistical model are also easier to interpret compared to other parametric models, because the splitting reveals causal relationships which are easy to understand and explain. For example, by analyzing the number of times each feature is used in a node to split data in a decision tree, we can understand the relative importance of different features and to determine those that are most influential for the predicted property\cite{breiman2017classification}.
But of all machine learning techniques, decision trees are amongst the most prone to overfitting because we cannot know \textit{a priori} how to formulate the smallest tree that completes the learning task, and all practical implementations must mitigate this challenge. This has led to specialized approaches such as pruning or bagging and boosting to prevent overfitting, as well as other regularization techniques also developed in deep learning such as early stopping and ensemble learning for which decision trees benefit from becoming "random forests"\cite{Goodfellow-et-al-2016}. Statistical learning models have been successfully applied to molecular property predictions, as in the example of modeling of different quantitative structure-activity relationships with a decision tree based on random forest optimization\cite{svetnik2003random}, and are starting to replace the use of SVMs in classification tasks more broadly.
\newline
\subsection*{\fontsize{11}{11}\selectfont Feature Representation}
\noindent
Similar to all modeling tasks, a representation or descriptor, is a mathematical abstraction of the inherent nature of the input, $\textbf{{x}}$, such as its chemical structure. Therefore, it is subject to the limitations of omitted features that may be influential for the property of interest. Thus, it is  common practice to add more physical details into the representation such that they then correlate better/easier with target properties, $\textbf{y}$.  In fact, research topics like quantitative structure property/activity relationships (QSPR/QSAR) have been popular and effective in the feature domain before modern machine learning has become more widespread. For ML, feature representations, when matched with the capabilities of the learning algorithms, are our most effective means to learn a chemical pattern/trend in data\cite{Haghighatlari2019}.

There are key criteria that we should consider for the construction of new descriptors:  
\begin{itemize}
\setlength\itemsep{0.1em}
	\item \textbf{uniqueness}. The representation should be unique with respect to the relative spatial arrangement of atoms. Often we need to develop descriptors that are invariant to the symmetries of the system (e.g., translation, rotation, atomic permutation, etc.), but are also distinctive for asymmetries (e.g., stereochemical chirality of molecules). Hence we prefer a one-to-one mapping not only for the easier training of ML models but also for a better generalizability and prediction performance.
	\item \textbf{universality}. The representation should be easily extendable to any system. If a descriptor is more representative to the fundamental chemical nature of the system, it also exhibits better transferability to new and future datasets. This is a key point for the accelerated exploration of molecular space, for example by means of virtual high-throughput screening\cite{Hachmann2018}. 
	\item \textbf{efficiency}. The representation should be computationally efficient. The key advantage of any ML model to its computational or experimental alternatives is the efficiency. However, for some type of descriptors the cost of feature representation is narrowly comparable to the generation of reference (computational) data. For example, this is specifically the case for higher-order many-body interactions \cite{Pozdnyakov2020}. 
\end{itemize}

Fulfilling all these criteria for the development of a desirable descriptor is a challenging task that necessitates expert  knowledge of chemistry and computer science. In addition, the comparison of descriptors in terms of performance and efficiency is a nontrivial task, as it strongly depends on the data type and molecular diversity. A fair comparison of feature representations requires the same training setup in terms of data set size and sampling and ML model complexity. The main reason is that if a data set is sparse and less representative of the entire molecular space, their feature representation is also limited to the available molecular makeup. Thereby, the resulting prediction performance is also restricted to the applicability domain of model that is imposed by training data.

Considering a broad spectrum of representations used to build ML models \cite{Faber2017a}, the required chemical information to encode molecular descriptors varies based on their availability and necessity for a given task. For example, inspired by QM we might consider atomic numbers, Z, and their chemical bonding sufficient to differentiate chemical systems from each other (2D descriptors). 
Moreover, if we aim to ultimately sidestep expensive QM calculations, we hope for the availability of atomic coordinates in order to correlate with rigorous electronic properties of the system (3D descriptors)\cite{Butler2018}. Basic inputs with topological features of chemical structures such as type and size of ring or walk and path counts are also useful.

The computational cost of obtaining the chemical information affects the overall efficiency of feature representation, and should be considered for their usage. For instance, the choice of 3D descriptors for training on QM computational data may require almost equally expensive geometry optimization for data generation. Thus, for future predictions the cost of preparing ML model inputs will be comparable to the reference QM calculations. However, if the atomic coordinates are available in advance, e.g., from experimental characterizations, or if the reference data is more demanding than geometry optimization (e.g., experimental data that is not easy to simulate such as melting point or solubility) the computational cost is often justifiable.

In addition, physicochemical properties such as electronegativity, polarizability, and ionization potential has been commonly used in the drug discovery community. These types of data can be obtained using first principles or data mining, and has its roots in the bioinformatics and cheminformatics domains. Descriptors based on such processed information are commonly referred to as hand-crafted descriptors or "engineered" features\cite{Rajan2013b}. 

The employed techniques for determining feature representation rely on different factors, including data type, ML algorithm, and of course the developer's creativity. For example, one may consider a molecule as a weighted graph with features assigned to its nodes and edges, i.e., atomic features and bond features, and their consecutive interactions of atomic and bond features of their nearest neighbours. Thus, the overall representation is built using local atomic environments that rely on 2D chemical information. In 2015, Duvenaud and coworkers applied this idea in the form of graph convolutional networks (GCNs) to generalize the well established fingerprint algorithms that describe molecular makeups\cite{Duvenaud2015}. The hierarchical complexity of GCNs helped to extract from the topological combination of atomic and bond features an accurate explanation of a variety of chemical properties. Since then, a large number of published studies have reported successful improvements by tuning types of atomic/bond features and their interactions\cite{Kearnes2016b, Yang2019}. Several recent studies also consider non-bonded interactions (i.e., disconnected nodes) by accounting for interatomic distances as pairwise features\cite{Schutt2018b, Chen2019b}(see Figure 2).

\begin{figure}[H]
\centering
\includegraphics[width=0.48\textwidth]{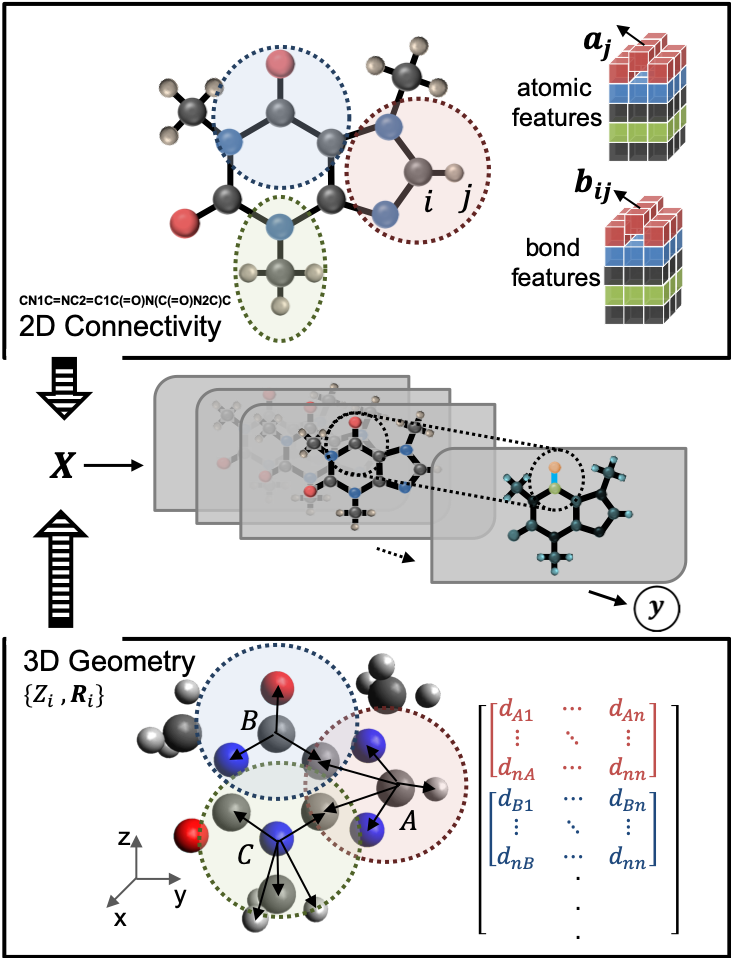}
\caption*{Figure 2. The illustration of graph convolutional networks with different representations of Caffeine molecule as input. Molecular information can be represented as atomic and bond feature tensors extracted from connectivity based 2D information, or as distance matrices obtained from 3D coordinates, or any other form of sensible representations.}
\end{figure}

Alternatively, one may consider many-body interactions beyond only pairs of atoms and assign a unique functional form, e.g. symmetry functions, to represent the histogram of available interactions up to a certain degree \cite{Behler2007, Bartok2013c, Christensen2020}. Thus, similar to composing molecular descriptors from atomic and pairwise features, they decompose many-body interactions and build a descriptor that relies on all terms individually and simultaneously.

More recent attention has focused on the provision of 3D structures with minimum information loss. The idea is to represent molecules to the ML model in the same way that they are visualized \cite{cnn_enzyme, 3d_densenet}, e.g., using a set of atomic densities. This type of representation has similarities with both QM, i.e., by providing electron density distribution of atoms, and computer vision, i.e., by replicating human vision using the complete configuration of the elements of a system \cite{comp_vision}.

Due to the flexibility in design and hierarchical manipulation of latent feature space, neural networks have become the cornerstone of creative ideas to integrate chemical information with the ML workflow.
The results of such efforts has created a new branch of feature representation that is commonly known as learned features.
Later in this paper, we present notable examples from our lab of employing engineered and learned features in the course of molecular property prediction.
\newline
\subsection*{\fontsize{10}{10}\selectfont Types of Data and their Abundance}
\label{subsec:data}
\noindent
The quality of labelled chemical datasets, composed of $(\textbf{x}, \textbf{y})$ pairs, is one of the key components in developing an accurate and predictive ML model. Even though generating systematic and exhaustive datasets which samples the chemical space computationally has arrived recently\cite{RN3044}, experimental datasets are indispensable because some properties are either difficult or impossible to compute. In developing ML models, one needs to be aware of the inherent differences between computational and experimental data, and take them into account when designing suitable representation for a given target property.

The reliability, accuracy, and reproducibility of computational data is improved by applying a concrete computational protocol across the dataset and carefully choosing and reporting its parameters, like level of theory, basis set, convergence criteria, and number of grid points. Even though similar standards can be applied in generating experimental data, the nature of experimental protocol or experimental conditions (e.g., solvent, temperature, pH) is most likely different as the data is commonly compiled from various sources. This leads to an inherent inconsistency in data compounded by different measurement errors in different experiments. For example, Nuclear magnetic resonance (NMR) chemical shift prediction utilizes X-ray crystal structures and solution NMR measurements to define the ($\textbf{x}, \textbf{y}$) labelled pairs, although their correspondence is not one-to-one.

The comprehensiveness of computational data is systematically improvable by continued enumeration of chemical compounds and their properties. In contrast, experimental measurements are time-consuming and resource-intensive, and adding additional data points to an experimental set is difficult, thus they sample chemical space more sparsely. This has led to combining experimental and computational data in some cases\cite{jha2019}. To further capture the inherent complexity of experimental data, their feature representation can be augmented with environmental conditions (like temperature, pH, and solvent). For example, hydrogen-bonding environments from crystal waters in the X-ray structure were also included in the prediction of a chemical shift of atoms in proteins to account for solvent effects\cite{ucbshift_arxiv}. Data augmentation from computation can be designed to incorporate ensemble averaging of experimental structures, such as introducing backbone flexibility commensurate with X-ray diffraction \cite{Friedland2010} and/or side chain repacking that reproduces NMR J-couplings\cite{Bhowmick2015} for proteins. Alternatively, one can include multiple input representations to the same property value which also increases the size of the dataset. Typically these augmentation approaches seek the sweet spot of low computational cost and high chemical/structural diversity to achieve the desirable experimental prediction accuracy. 
\newline
\section*{\fontsize{10}{10}\selectfont INTERPLAY OF REPRESENTATION, DATA, AND MACHINE LEARNING ALGORITHM TO PREDICT CHEMICAL PROPERTIES}
\label{sec:nmr}
\noindent
NMR spectroscopy is one of the most important molecular probes of chemical composition, structure and dynamics of small molecules through to large proteins. The least invasive techniques of NMR are the chemical shifts and spin-spin splittings which can be measured to very high accuracy. Because they are sensitive to their functional groups, detailed geometries, and chemical environments, they allow for prediction of solution phase protein structures or to identify or verify the structure of chemical compounds in the crystalline phase.\cite{nmr_xrd}

The connection between NMR chemical shifts to structural or dynamical properties, while true in principle, is nevertheless sometimes difficult to reveal in practice through direct assignment of the spectrum. One solution to this problem is to rely on expensive QM methods that often can accurately predict spectral observables from structure of small molecular fragments\cite{gipaw}. While chemical physics approaches have achieved considerable success in spectral assignment and structure determination, here we consider two recent examples of supervised learning approaches where the interplay of chemical descriptors, data size and augmentation strategies, and choice of ML algorithm has significantly improved the accuracy of chemical shift predictions and their connections to complex structure in aqueous solution and in the solid state.
\newline
\subsection*{\fontsize{10}{10}\selectfont Engineered Features and Random Forest Regression to Predict Chemical Shifts for Aqueous Proteins}
\label{subsec:nmr_protein}
\noindent
Given the expense of QM calculation for magnetic properties, heuristic NMR "calculators" have been developed for efficient chemical shift evaluations for aqueous proteins. In particular, the single-layer feed-forward network developed and packaged as SPARTA+\cite{sparta+} remains among the most popular of chemical shift prediction methods. Better predictive power can also be gained by exploiting sequence homology as that used in SHIFTX2\cite{shiftx2}, as the expectation is that as more sequence and spectroscopic data is deposited in public repositories, it will allow interpolation to replace extrapolation for a variety of NMR observables. 

Even with these successes, these algorithms are still open to change as modern day ML approaches march forward alongside accumulating biological data. Furthermore, engineered features are ideal for predicting experimental chemical shifts in solution because they better model the averaged chemical shifts over different instantaneous conformations of the structure in the thermalized ensemble. Therefore classification features, like whether an atom is involved in a hydrogen bond or a residues secondary structure category, are relatively stable for different conformations in the ensemble relative to the coordinates of the atoms in 3D space.

Recently engineered features extracted from protein X-ray crystal structures has been utilized together with random forest regression to formulate the UCBShift chemical shift predictor for aqueous proteins.\cite{ucbshift_arxiv} All backbone atoms and the side chain $\beta$-carbon chemical shifts of a residue are mapped from numerical and non-numerical features built from the geometries and biophysical properties of a tripeptide centered at the target residue. The features include backbone and side chain torsion angles, BLOSUM numbers identifying the likelihood of residue substitution, secondary structure, hydrogen bond geometries, ring currents, half surface exposure, accessible surface area, and non-linear transformations of distance features which have physical relevance from QM. All of these features are formulated as internal properties of the protein which naturally exhibit translational and rotational invariance. 

\begin{figure*}[t]
\center
\includegraphics[width=0.75\textwidth]{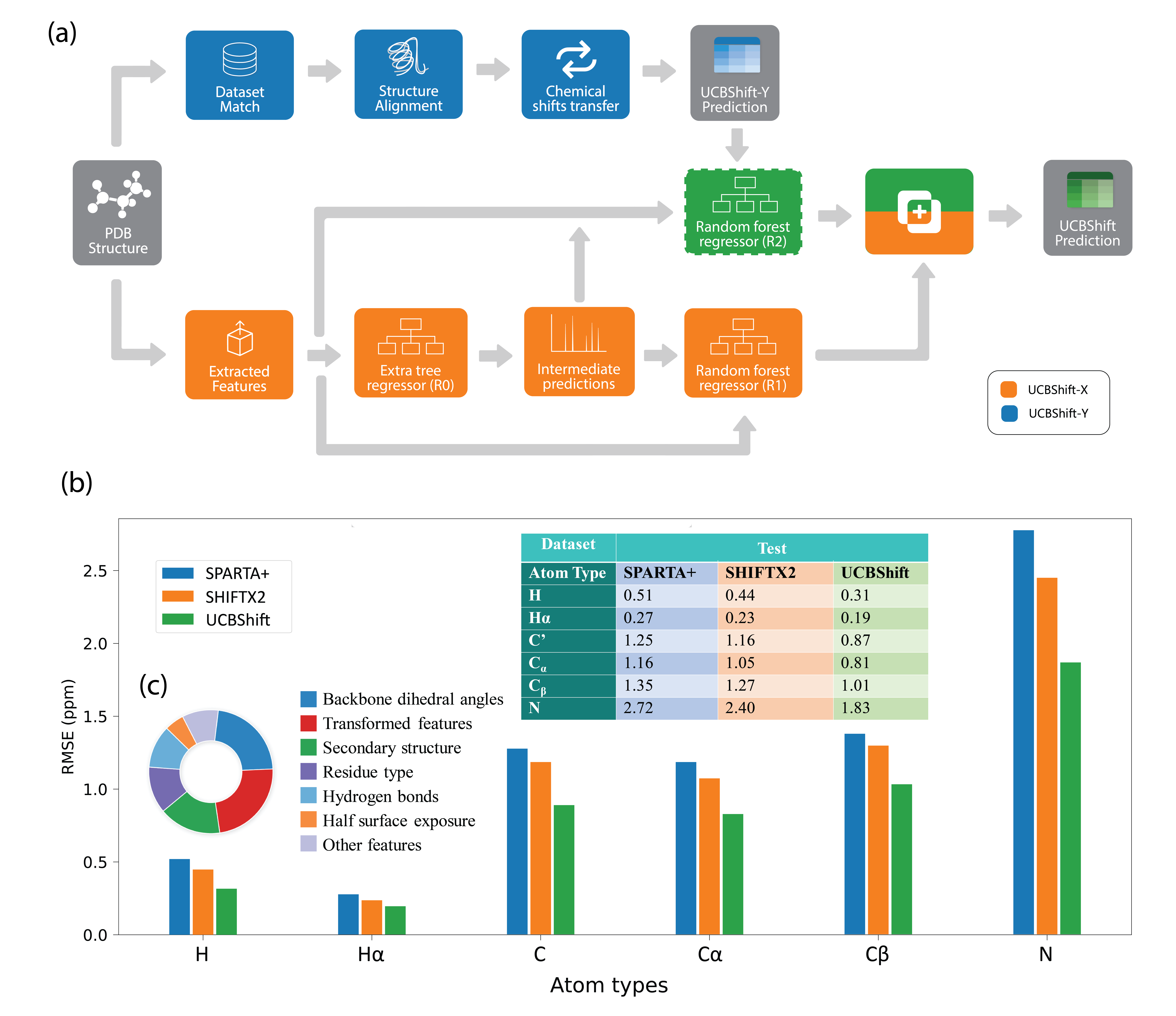}
\caption*{Figure 3. (a) Illustration of the UCBShift algorithm\cite{ucbshift_arxiv} (b) Testing RMSEs (ppm) for each atom type from SPARTA+, SHIFTX2 and UCBShift, when evaluated on an independently generated test dataset (c) Relative importance of all the input features analyzed from UCBShift model.}
\label{fig:Jerry}
\end{figure*}

However, the universality of the representation is limited to proteins without functional group modifications or bonding with ions, ligands or other hydrogen-bonding motifs with water. To increase the applicability of our ML model, we have also included extraction of crystal water positions in the evaluation of features such as hydrogen bonding, and alignment scores that characterize sequence and structural homology to other proteins with recorded chemical shifts, aiding the chemical shift prediction through learned direct transfer if the similarity is faithful enough to the query protein.\cite{ucbshift_arxiv}  

The UCBShift algorithm utilizes two successive decision tree ensemble models (Figure 3a), one which differentiates the various atomic environments in a protein utilizing engineered features, and a second that make predictions based on the most similar sequence and/or structural alignments in the training dataset. As a result, UCBShift has significantly lower root-mean-square-error (RMSE) when applied to an independently generated test dataset when compared to SPARTA+ and SHIFTX2 on all the relevant protein atom types (Figure 3b). Further analysis of the total number of decisions made in each tree, which is visualized in Figure 3c, reveals that the  QM-inspired transformations of the features account for more than 20\% of the feature importance.\cite{ucbshift_arxiv}

\begin{figure*}[ht]
\center
\includegraphics[width=0.75\textwidth]{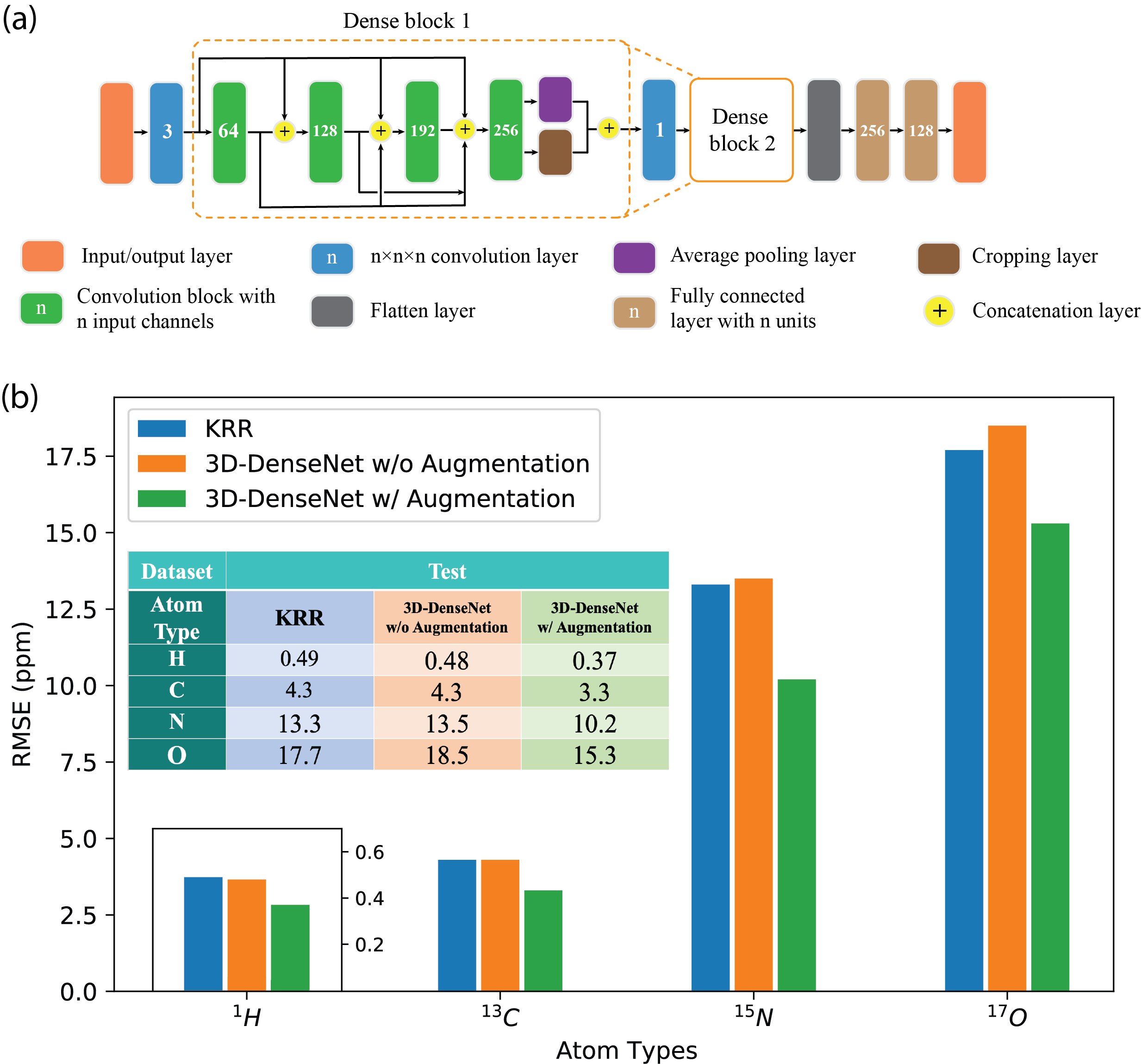}
\caption*{Figure 4. (a) Illustration of MR-3D-DenseNet architecture (b) Testing RMSEs (ppm) for each atom type from KRR, 3D-DenseNet without data augmentation and 3D-DenseNet with data augmentation.}
\label{fig:shuai_model}
\end{figure*}

SPARTA+\cite{sparta+} and SHIFTX2\cite{shiftx2}, which are based on simpler machine learning models, as well as our own attempts with deep recurrent neural networks with residual connections, have not performed as well as the random forest model presented here. This is because simple MLs do not have sufficient capacity to recognize the complexity of the mapping from engineered features to chemical shifts, and the limited number of well-formulated structure-chemical shift pair in the dataset prevents those more complicated deep neural networks to effectively train. This consequence once again reinforces the importance to wisely choose a ML algorithm depending on the size of the dataset and a featurized representation of the molecular system to realize the excellent predictive power for solution-phase NMR properties.
\newline
\subsection*{\fontsize{10}{10}\selectfont Chemical Shift Prediction in the Solid State using DL Architectures and Data Augmentation}
\label{subsec:nmr_solid}
\noindent
Crystal structures of small molecules can be identified by comparing the experimental measurements of solid-state NMR chemical shifts with the calculated results using DFT, typically using the Gauge-Including Projector-Augmented Waves (GIPAW) method \cite{gipaw}. However, because of the cubic scaling with the size of the atomic basis sets used in the DFT calculation, ML algorithms have been investigated to approximate the QM physics. For example,  a shallow ANN using engineered features was used to predict chemical shifts (and quadrupolar couplings) in silica materials using symmetry functions operating on the Cartesian coordinates to respect rotational invariance of the chemical shift value to applied magnetic field\cite{Cuny2016}. Paluzzo et al. devised a ML approach using 3D structures, while also directly incorporating rotational symmetry using KRR and the SOAP kernel\cite{Bartok2013c}, yielding very good results for chemical shift prediction for small molecule crystal systems\cite{ml_nmr2}. Even though a significant acceleration factor was achieved over QM using these ML approaches, the training data generation using DFT is itself a bottleneck, thereby making a shallow ANN necessary, while the quadratic-to-cubic complexity for calculating and inverting the kernel matrix makes it also impractical for KRR to treat larger datasets. 

The question we set out to address was whether a deep learning approach was tenable for the prediction of DFT chemical shifts for hydrogen ($^1$H), carbon ($^{13}$C), nitrogen ($^{15}$N) and oxygen ($^{17}$O) of organic molecules in molecular crystals. The input representation was comprised of the 3D coordinates of atoms in the unit cell taken from the Cambridge Structural Database (CSD), "imagery" that was ideally suited to a multi-resolution CNN based on a DenseNet approach as shown in Figure 4a. 

We utilized the chemical environment for each atom whose chemical shift is predicted is represented on a 3D grid with a calculated Gaussian density at each atom center. This input representation describes local bonding characteristics
that arrange atoms into 3D shapes with more global spatial organization. CNNs are ideally suited to the 3D structural data and electron density representation. This is because the network architecture of a CNN was originally formulated to operate on data that has temporal organization, i.e. 2D images arranged in a time series, but for which the time axis can be replaced by a 3rd spatial dimension to represent the electron density distribution. Hence  We benefited from the open access to the original DFT chemical shifts calculated on 2000 organic molecules containing $\sim\ 30-40$ atoms to create the labelled data.\cite{ml_nmr2} 

Furthermore, better data representations and data quantity proved crucial to the success of our DL approach. First, we showed that the chemical environment for each atom type could be represented by multiple resolutions (MR), thereby incorporating the atomic densities of the other atoms over different grid sizes of $d$ (4\AA, 6\AA, 8\AA, 10\AA, and 14\AA) with $16 \times 16 \times 16$ voxels, and representing each resolution with its own dedicated channel. Under each resolution, we divided the density based on the atom types into 4 different channels for $^1$H, $^{13}$C, $^{15}$N, $^{17}$O, respectively, similar to RGB channels used in image recognition. Second, given the limited number of examples in the training dataset, and the prohibitive expense of creating an order of magnitude more data, we recognized that a cheap data augmentation method was obviously available. Instead of enforcing chemical shift  invariance through explicit rotational symmetry operations, we instead just augmented the data by rotating the Cartesian coordinates of atoms randomly with the Euler angles uniformly distributed between [$-\frac{\pi}{2}, \frac{\pi}{2}$] along each of $x$, $y$ and $z$ axis. During the training phase, both the original data and augmented data are included in the training dataset, while during the testing phase we average the prediction results over the different rotation configurations, thereby manifesting ensemble learning.\cite{Goodfellow-et-al-2016} 

Figure 4b summarizes the results for the  MR-3D-DenseNet workflow in terms of the root mean square error (RMSE) for chemical shift prediction of the DFT results for each atom type. Using the greater capacity of the MR-3D-DenseNet deep network, we obtain nearly 15\%
improvement for $^{17}$O and close to ~25\% for $^{13}$C, $^{15}$N, and $^1$H chemical shifts over KRR, for which hydrogen chemical shifts are similar in error between \textit{ab initio} calculations and experimental measurements. 
\newline
\section*{\fontsize{10}{10}\selectfont CONCLUSIONS AND OUTLOOK}
\label{sec:conclusion}
\noindent
We have shown that three key components of a machine learning workflow - algorithm, features, and data - are inexorably intertwined for achieving predictive success for biological, chemical, and materials applications. Inconsistent decisions about choice of ML algorithms and feature representation with respect to size and source of data can lead to inaccurate deployment of techniques in molecular property prediction.
We illustrated that proper execution of the ML triad can lead to successful prediction of NMR chemical shifts of molecules in solution or for crystalline states. 

Not surprisingly, these three key components are also at the heart of current developments in ML, and many open questions and challenges need to be addressed to push the boundaries toward new applications.
In terms of feature representation, systematic development of new descriptors, standardization of their evaluation, and easier accessibility via user interfaces (e.g., Python libraries) are necessary to establish their long-term development \cite{chemml}. A transparent and sustained study of feature representation would involve researchers with variety of domain knowledge and expertise to accelerate future developments.  

In addition, technical challenges involved with scarce and sparse data sets need to be vigorously discussed, as it is often a prevalent case for applying ML to real-world chemical applications when expensive calculations or difficult experiments are the bottleneck. The good news is that the entire ML community has been giving more attention to this issue, and  
as a result, techniques that can deal with limited data have been growing: 
\begin{itemize}
\setlength\itemsep{0.1em}
\item methods that are intrinsically suitable for small size data, such as kernel methods or low-variance models with feature reduction capabilities,
\item methods that leverage small size data in the learning task, for example by transfer learning \cite{Yamada2019} from pre-trained or high-fidelity models, or in some cases by multitask learning,
\item methods for data curation, such as learning to impute missing data or by physically meaningful data augmentation\cite{Schwaller2019, 3d_densenet}
\item decreasing the number of data generation trials via sampling methods using an  active learning (AL) approach \cite{Sifain2018, chemml}; AL methods can discover the uncertainty of trained models in the high-dimensional data distribution and query more informative training data that improves the model most
\item future work would be to investigate imbalanced data and underrepresented regions of the solution space studied in the form of unsupervised ML techniques, such as clustering methods.
\end{itemize}

Moreover, future research needs to examine more closely the \textit{interpretability} of chemical ML models. In this regard, gaining chemical insights and understanding direct relationships between molecular observables and their properties is the ultimate goal. The issue with model interpretability is that ML algorithms are designed to learn patterns (or mappings) in high-dimensional data that is otherwise not obvious to ourselves. Thus, predictive models are generally perceived as a so-called black-box model with limited transparency\cite{Roscher2019}. However, we argue that this would not be the case if we work with hand-crafted features and the simplest possible ML model. Hence part of the current concern regarding interpretability of ML models stems from highly parameterized models with arbitrary choice of hidden states. Thus, a general practice for the future work is to simplify state-of-the-art models and evaluate model shrinkage as part of a cost-benefit analysis. For instance, a learning curve based on the number of trainable parameters (or number and type of layers) should become a trend in applications of deep learning models.

On the other hand, ML complexity is still sometimes needed, and the interpretability of models requires greater contributions from expertise equipped with domain knowledge. For example relevance propagation techniques\cite{Bach2015} can help interpret a trained ML model. Recent efforts on developing visualization tools can also help to monitor the change/gain by adding extra layers to deep learning models.

Often there is a fear that a new approach such as ML could supplant existing computational chemistry techniques or suppress models designed for physical insight. This will never be the case.  Given the considerable progress in algorithms, molecular feature representation, and data accessibility, we expect interest in applying ML to almost any vein of chemical and materials sciences will continue to grow and ultimately settle in as a long term player in the general computational chemistry landscape.
\newline
\section*{ACKNOWLEDGMENTS}
\noindent
The authors thank the National Institutes of Health for support under Grant No 5U01GM121667. FHZ thanks the Research Foundation-Flanders (FWO) Postdoctoral Fellowship for support of this work. XG thanks the U.S. DOE under the Basic Energy Sciences CPIMS program, Contract No. DE-AC02-05CH11231 for research support of ML to chemical reactions. This research used computational resources of the National Energy Research Scientific Computing Center, a DOE Office of Science User Facility supported by the Office of Science of the U.S. Department of Energy under Contract No. DE-AC02-05CH11231.
\newline
\section*{REFERENCES}
\bibliographystyle{unsrt}
\bibliography{networks}

\end{document}